# Kinetics of Laser-Assisted Carbon Nanotube Growth

Y. van de Burgt,[a,b] Y. Bellouard[a] and R. Mandamparambil[a,b] ,

Laser-assisted chemical vapour deposition (CVD) growth is an attractive mask-less process for growing locally aligned carbon nanotubes (CNTs) in selected places on temperature sensitive substrates. The nature of the localized process results in fast carbon nanotube growth with high experimental throughput. Here, we report on detailed investigation of growth kinetics related to physical and chemical process characteristics. Specifically, the growth kinetics is investigated by monitoring the dynamical changes of reflected laser beam intensity during growth. Benefiting from the fast growth and high experimental throughput, we investigate a wide range of experimental conditions and propose several growth regimes. Rate-limiting steps are determined using rate equations linked to the proposed growth regimes, which are further characterized by Raman spectroscopy and Scanning Electron Microscopy (SEM), therefore directly linking growth regimes to the structural quality of the CNTs. Activation energies for the different regimes are found in the range of 0.3 – 0.8 eV.

## Introduction

Vertically aligned carbon nanotube (CNT) structures are desirable for many applications. Their unique structural, electronic and thermal properties make them ideal candidates for applications such as field emitters,[1,2] field effect transistors,[3,4] filters,[5–7] interconnects[8] and sensors.[9] So far, producing aligned nanotube structures has been obtained almost exclusively by chemical vapour deposition (CVD). This process allows a high degree of control over the resulting growth and morphology by tailoring the catalyst that adsorbs and dissociates the carbonaceous gas. Despite enhanced control, the quality of CVD-produced nanotubes is generally low as a result of a relatively high defect presence. Further improvements can be achieved through a better understanding of the growth kinetics and the various catalytic mechanisms.

Carbon nanotube growth kinetics and its dependence on process gases are currently not fully understood. The precise reaction process is still under debate.[10] Growth-limiting reaction steps depend on precise growth conditions and resulting activation energies vary widely between experimental conditions such as temperature, pressure, catalyst and gas composition but also between subsequent experiments with seemingly identical conditions.

A promising, fast method for local growth of aligned carbon nanotubes is laser-assisted CVD growth, in which a laser locally heats a substrate.[11,12] Since the heating is much faster compared to conventional thermal CVD, an in situ preparation of the catalyst is possible as opposed to introducing an extra catalyst reduction step.

We use a small reaction chamber with precise control of the composition of the laminar flow of process gases, as described in previous work.[13] This allows for *in situ* determination of the growth kinetics while maintaining a higher experimental throughput than conventional CVD growth. Specifically, we investigate the influence of feedstock gas flow and composition as well as the corresponding kinetics involved in laser-assisted CVD growth of carbon nanotubes. From kinetic relations, we derive and propose several different growth regimes and corresponding rate constants. The regimes are linked to elementary steps of the CNT growth process, as schematically proposed in Fig 1. The influence of the three different process gases, ethylene, argon and hydrogen, is related to the proposed regimes as confirmed by micro-Raman spectroscopy quality measurements. Using time-resolved reflectivity measurements, first order dynamics of the growth of carbon nanotubes are obtained which are used to calculate activation energies for the different regimes. Finally, by exploring the growth rate of the carbon nanotubes as a function of several partial pressures we derive the various reaction orders.

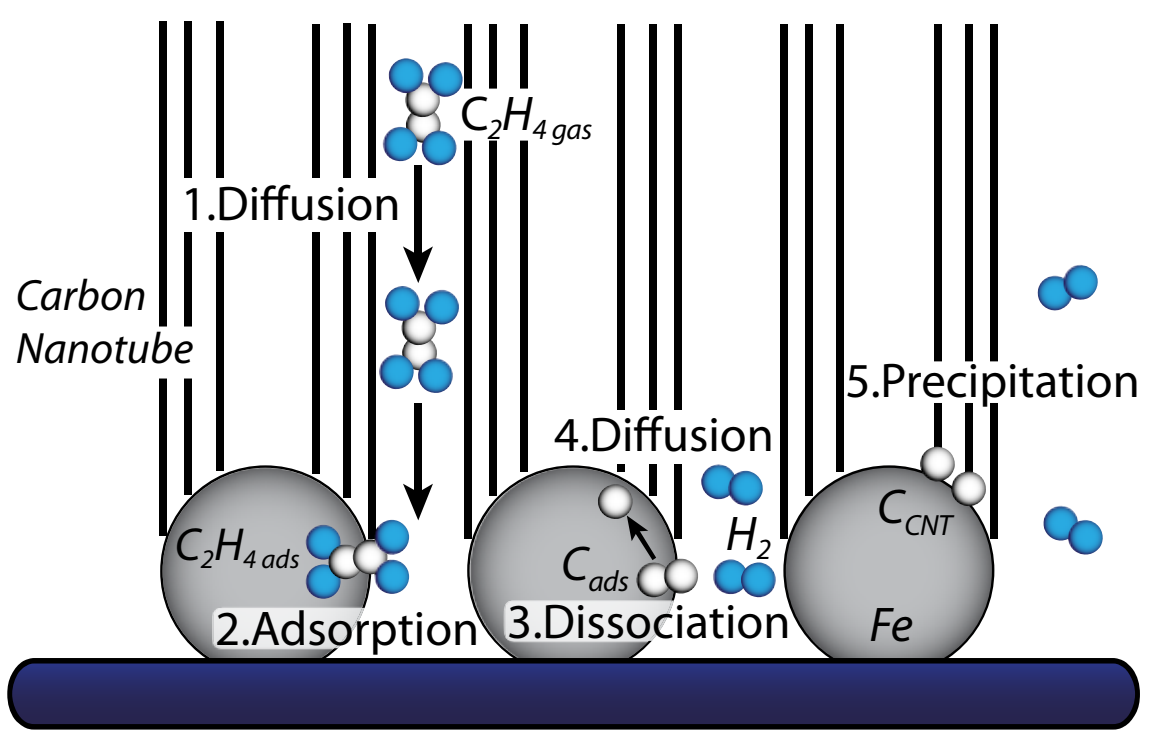

Fig 1. Schematic of the carbon nanotube growth process investigated in this study. 1. Diffusion of the ethylene through the CNT forest. 2. Adsorption of ethylene onto the catalyst. 3. Dissociation of ethylene into adsorbed carbon and hydrogen gas. 4. Diffusion through the catalyst. 5. Growth of the carbon nanotube.

[a] Department of Mechanical Engineering, Eindhoven University of Technology, Den Dolech 2, Eindhoven, The Netherlands. E-mail: y.b.v.d.burgt@tue.nl

[b] Holst Centre/TNO – Netherlands Organization for Applied Scientific Research, HTC31, Eindhoven, The Netherlands.

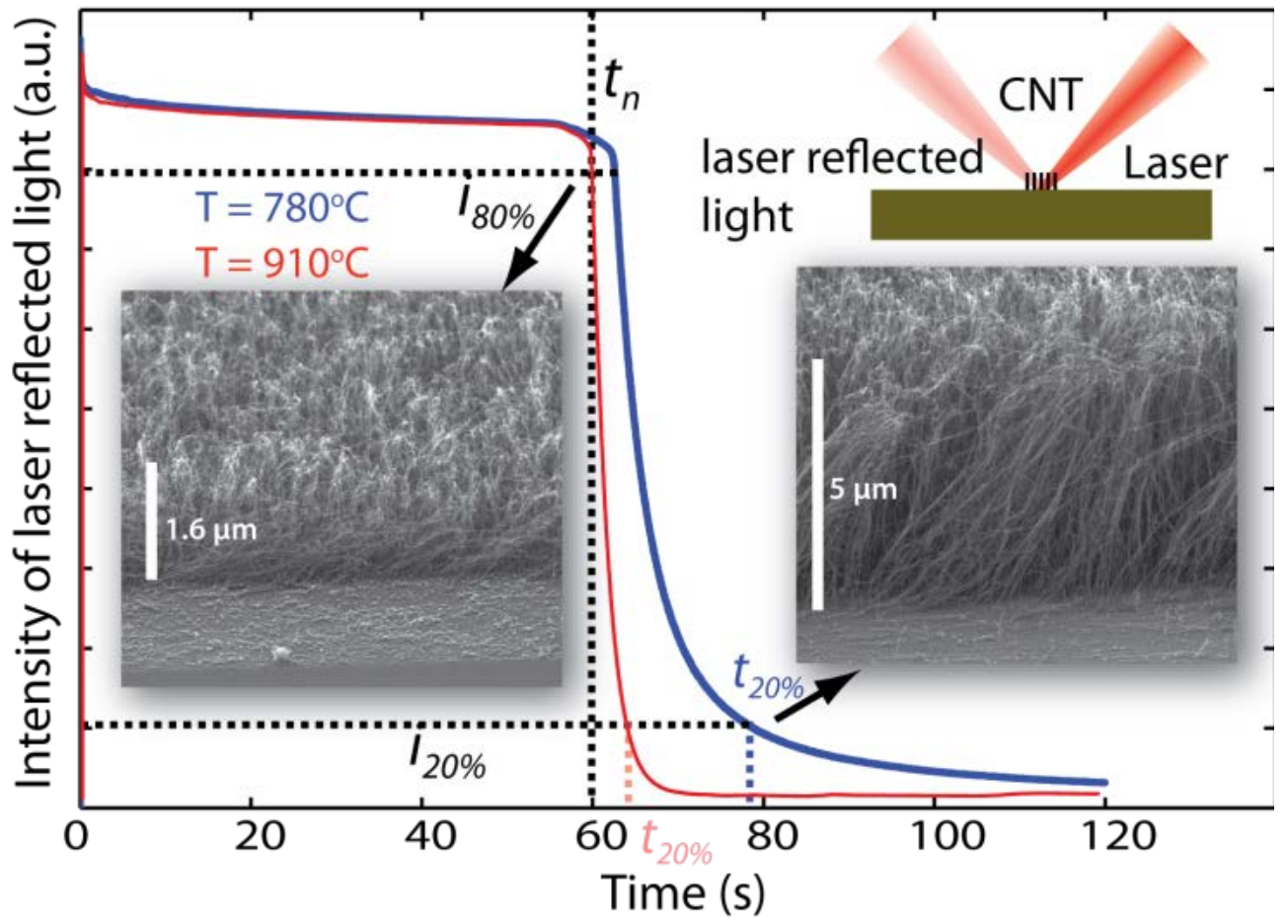

Fig 2. Plot of the intensity of reflected laser signal versus time for two different experiments. Indicated in the figure are the nucleation time tn and the intensity level of the signal of 80% and 20% of the maximum value, i80% and i20% respectively. For the 20% intensity the time at which this is reached is depicted for both experiments as well. Also shown are the corresponding SEM pictures of the nanotube forest at the intensity levels 80% and 20%. The inset shows a schematic representation of the growth of CNTs by laser irradiance and the laser reflected signal.

The laser-assisted growth method investigated in this paper is of specific interest for determining carbon nanotube growth kinetics in general, as the nature of the localized laser-heated process allows for a fast and high throughput *in situ* investigation of the growth characteristics with respect to the experimental conditions.

## Experimental and Theoretical Methods

The experimental setup consists of a feedback-controlled laser-assisted carbon nanotube growth process.[14] The particular design of the reaction chamber allows the process gases (argon, hydrogen and ethylene) to flow in a controlled manner over the substrate. The setup is further described in Methods. To investigate growth kinetics, the reflected laser signal is dynamically measured. A sudden drop in reflection is attributed to the initiation and growth of the carbon nanotube forest,[15] indicated with $t_n$ in Fig 2. The average growth rate is calculated by dividing the estimated length of the nanotube forest by the growth time ($t_{20\%} - t_n$) as indicated for two cases in the figure. This is done for several process conditions to confirm the CNT length for a specific intensity. The time $t_{20\%}$ is chosen to be shorter than the lifetime of the catalyst since the CNT growth continues, demonstrated by the exponential decay of the reflected laser intensity. Combined with FEM modelling, the temperatures involved in an experiment can be calculated which allows us to investigate activation energies. See Methods for a more detailed description.

The decay of the intensity seems to follow the Beer-Lambert law,

$$I(z,t) = I_0 e^{-2\alpha \overline{z(t)}} \tag{1}$$

where $I$ is the intensity, $z$ is the average thickness of the layer and $\alpha$ its absorbance. In that case the absorbance $\alpha$ can be estimated with,

$$\alpha = \frac{\ln\left(I_0 / I\right)}{2z} \tag{2}$$

Several lengths of CNTs and corresponding intensities were used to estimate $\alpha$. Surprisingly, this resulted in a variation of the absorbance over time in contrast to results from Puretzky *et al.*[16] This dynamically changing absorbance might imply a CNT density variation during growth. For that reason, we manually measure the CNT forest lengths at their highest point under different experimental conditions. This method bypasses the uncertainties in the growth rate calculations as a result of the varying absorbance $\alpha$ and enables us to estimate the average growth rate.

The catalytic chemical vapour deposition growth of carbon nanotubes can be schematically represented by a number of elementary steps.[16,17] Those steps are graphically presented in Fig 1. The ethylene gas has to reach the catalyst, by a forced flow or natural diffusion. Diffusing through the nanotube forest might play a role, although that process is assumed to be very fast; giving notable effects only at nanotube lengths much greater (in the order of mm) than what we expect here.[18] When the ethylene reaches the catalyst, it is adsorbed onto the iron catalyst surface. This step is followed by a dissociation step, resulting in adsorbed carbon and hydrogen. The adsorbed carbon diffuses through or over the catalyst to the growth side where it precipitates as part of the carbon nanotube.

The following relations can be given regarding the growth kinetics of the different elementary steps,

$$C_2H_4 \underset{k_0^-}{\overset{k_0^+}{\rightleftharpoons}} C_2H_{4\,\mathrm{ads}} \tag{3}$$

$$C_2H_{4\,\mathrm{ads}} \underset{k_1^-}{\overset{k_1^+}{\rightleftharpoons}} 2C_{\mathrm{ads}} + 2H_2 \tag{4}$$

$$C_{\mathrm{ads}} \xrightarrow{k_2^+} C_{\mathrm{CNT}} \tag{5}$$

Here $k^+$ and $k^-$ are the rate constants for the forward and backward reaction, for respectively the reaction from ethylene gas into adsorbed ethylene $k_0$, from adsorbed ethylene into adsorbed carbon and hydrogen $k_1$, and from adsorbed carbon into carbon nanotube $k_2$.

We assume equilibrium of gas phase ethylene with adsorbed ethylene under normal process conditions. When we further assume that the gas phase concentration of ethylene does not change, *i.e.* a sufficient supply of ethylene gas is present, Equation (3) becomes,

$$\frac{\mathrm{d}\left[\mathrm{C_2H_4}\right]}{\mathrm{d}t}=0, \quad \rightarrow K_0=\frac{k_0^+}{k_0^-}=\frac{\left[\mathrm{C_2H_{4\,ads}}\right]}{\left[\mathrm{C_2H_4}\right]} \qquad (6)$$

where $K_0$ is an equilibrium constant for the reaction between ethylene and adsorbed ethylene. From the theory of kinetics and Equation (4), we can write the rate equation for the adsorbed carbon as a function of the concentration (or partial pressure) of the different components,

$$\frac{\mathrm{d}\left[\mathrm{C_{ads}}\right]}{\mathrm{d}t}=2k_1^+\left[\mathrm{C_2H_{4\,ads}}\right]-\left(k_2^++k_1^-\right)\left[\mathrm{C_{ads}}\right]\left[\mathrm{H_2}\right] \qquad (7)$$

Note that for a total pressure of 1 bar, the concentration and partial pressure of the different gases are interchangeable in these relations. We assume that the hydrogen concentration is constant, *i.e.* the produced hydrogen from dissociation is much less than the feedstock hydrogen. In other words, we assume,

$$\frac{\mathrm{d}[\mathrm{H_2}]}{\mathrm{dt}}=0, \quad \left[\mathrm{H_{2\,diss}}\right]<<\left[\mathrm{H_{2\,feed}}\right] \qquad (8)$$

Solving the linear differential equation (7) results in

$$\left[\mathrm{C_{ads}}\right]=\frac{C}{e^{(k_2^++k_1^-)[\mathrm{H_2}]t}}+\frac{2k_1^+}{k_2^++k_1^-}\frac{\left[\mathrm{C_2H_{4\,ads}}\right]}{\left[\mathrm{H_2}\right]} \qquad (9)$$

where $C$ is an integration constant. If we then assume no carbon is adsorbed at $t = 0$ this can be further reduced to

$$\left[\mathrm{C_{ads}}\right]=\frac{2k_1^+}{k_2^++k_1^-}\frac{\left[\mathrm{C_2H_{4\,ads}}\right]}{\left[\mathrm{H_2}\right]}\left(1-e^{-k_2^+[\mathrm{H_2}]t}\right) \qquad (10)$$

In this expression the exponential term defines the “start-up” phase, which is depending on the hydrogen partial pressure. In steady-state, *i.e.* for sufficiently large $t$, the exponential term disappears. The growth rate, $R_G$, of the carbon nanotubes can be expressed as,

$$R_G=\frac{\mathrm{d}\left[\mathrm{C_{CNT}}\right]}{\mathrm{d}t}=k_2^+\left[\mathrm{C_{ads}}\right] \qquad (11)$$

Taking into account all elementary steps depicted in Fig 1, we identify different growth regimes having specific rate-limiting steps that arise for different process conditions. It is generally assumed that the rate-determining step is equal to the process step with the highest energy barrier. However, different experimental conditions can lead to different energy barriers and corresponding growth regimes. This can be explained by the fact that energy barriers are usually a combination of different elementary steps, which can act differently under different conditions such as temperature and partial pressure but also coverage portion of the catalyst. The resulting activation energies have been found to vary widely in literature depending on the rate-limiting mechanism. The rate-limiting growth regimes for the catalytic CVD growth of carbon nanotubes that are considered here are *adsorption-limited,*[19] *surface diffusion-limited,*[19,20] *dissociation-limited,*[21,22] and *mass diffusion-limited*[23,24] and are schematically depicted in Fig 3. By varying the experimental conditions, the characteristics of these regimes are investigated.

**Adsorption-limited regime**

It was shown by Lebedeva *et al.*[19] that for low ethylene pressures, the adsorption of ethylene on the catalyst surface is the rate determining step. Although in our system a carrier gas is also present, this inert gas does *not* participate in the chemical reaction and as such we assume this regime to be equivalent to the one described in reference[19]. The effective coverage of ethylene on the catalyst is very low and the dissociation step, $k_0^+$, becomes the pre-dominant step. This is schematically shown in Fig 3(a) where the red coloured rate constant $k$ indicates the rate determining step. For this, we assume that the dissociation step from adsorbed ethylene to adsorbed carbon, $k_1^+$ is much faster. Furthermore, $k_0^+$ is lower than the surface diffusion and precipitation step, $k_2^+$. This will result in low growth rates and, without sufficient carbon supply, the introduction of defects.

For a constant pressure of 1 bar, this regime only occurs with a high partial pressure of argon gas and corresponding high flow rate. These high flow rates ensure a steady-state ethylene concentration around the catalyst and as a result the local gas phase ethylene concentration and partial pressure do not change.

**Surface diffusion-limited regime.**

With increasing ethylene partial pressure, sufficient ethylene can reach the surface of the catalyst and can reach equilibrium with the adsorbed ethylene and carbon. The surface diffusion and nanotube precipitation become the rate-determining step, where the reaction rate of the growth $k_2^+$ is much lower than the rate of reaction back to ethylene, $k_1^-$, see Fig 3(b). This means the first step is virtually at equilibrium and we can assume a steady state of the adsorbed carbon; in which the carbon nanotube growth does not affect the concentration of the adsorbed carbon

$$\frac{\mathrm{d}\left[\mathrm{C_{ads}}\right]}{\mathrm{d}t}=0, \quad k_2^+<<k_1^- \qquad (12)$$

We can therefore further simplify Equation (7) and write,

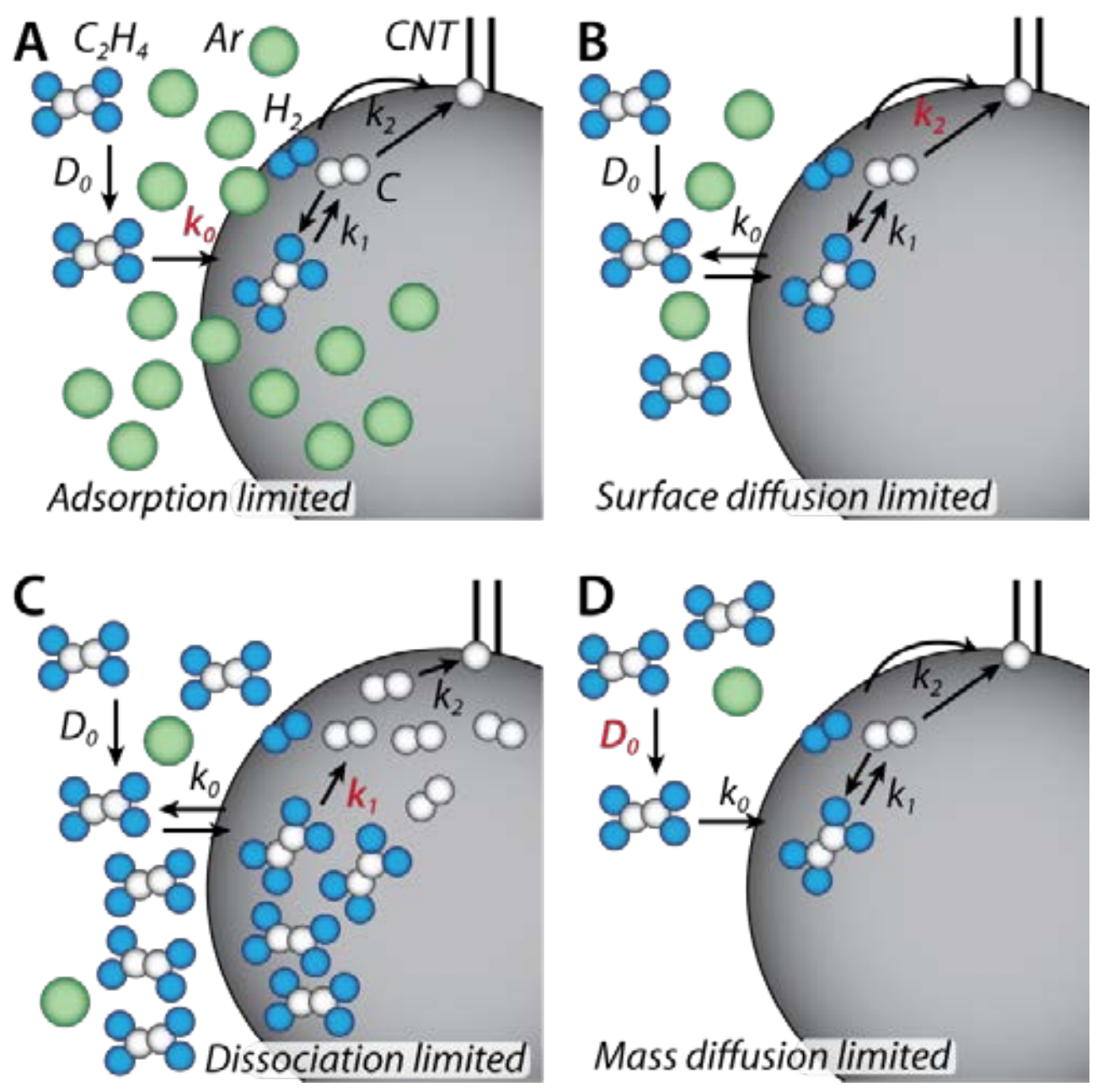

Fig 3. Schematic overview of different catalytic carbon nanotube growth regimes. A. Adsorption-limited: For low ethylene partial pressure, the adsorption on the catalyst is the rate-limiting step. B. Surface diffusion limited regime: diffusion through the catalyst and precipitation of the nanotube are expected to be the rate-limiting step. C. Dissociation limited regime: the dissociation of ethylene into carbon is considered the rate-limiting step. D. Mass diffusion limited: for low velocities, the renewal rate is low which results in local under-supply of ethylene.

$$K_1 = \frac{2k_1^+}{k_1^-} = \frac{[\mathrm{C_{ads}}][\mathrm{H_2}]}{[\mathrm{C_2H_{4\,ads}}]} \quad (13)$$

$$[\mathrm{C_{ads}}] = K_1 K_0 \frac{[\mathrm{C_2H_4}]}{[\mathrm{H_2}]} \quad (14)$$

where $K_1$ is an equilibrium constant and $K_0$ is from Equation (6) . Using Equation (11) we can then write

$$R_G = k_2^+[\mathrm{C_{ads}}] = k_2^+ K_1 K_0 \frac{[\mathrm{C_2H_4}]}{[\mathrm{H_2}]} \quad (15)$$

From this relation we note that the growth rate in this regime is depending on both the ethylene and hydrogen concentration. The growth rate is also proportional to the equilibrium constants $K_0$ *and* $K_1$ of the reaction between gas phase ethylene and adsorbed ethylene as well as between adsorbed ethylene and carbon and hydrogen, respectively.

**Dissociation-limited regime**

A further increase of the ethylene partial pressure causes the equilibrium of adsorbed ethylene and carbon to shift towards a nearly complete covering of adsorbed ethylene on the catalyst. In this case any dissociated ethylene will almost immediately be incorporated into a nanotube. Typical for the dissociation-limited regime is that the dissociation rate becomes the rate-limiting step. In this case, the rate-determining chemical equation is,

$$\mathrm{C_2H_{4\,ads}} \underset{k_1^-}{\overset{k_1^+}{\rightleftarrows}} 2\mathrm{C_{ads}} + 2\mathrm{H_2} \quad (16)$$

Because the dissociation transition from adsorbed ethylene into adsorbed carbon is the rate-determining step, we can assume the process of diffusion and incorporation of the carbon into a nanotube is faster and thus $k_1^-$ is assumed to be negligible. In fact, a very high partial pressure of ethylene could cover the whole catalytic active surface. The saturated catalyst could prevent the reaction back into adsorbed ethylene; at the same time a large surface coverage could result in non-linear behaviour with respect to the ethylene partial pressure. This is schematically represented in Fig 3(c).

Overall, this reaction is limited by the reaction rate constant $k_1^+$ and as such the concentration of adsorbed ethylene is assumed to be in equilibrium with the gas phase ethylene via the equilibrium constant $K_0$. Using Equation (10) we can write for the steady-state situation,

$$[\mathrm{C_{ads}}] = \frac{2k_1^+}{k_2^+}\frac{[\mathrm{C_2H_{4\,ads}}]}{[\mathrm{H_2}]} \quad (17)$$

where we have neglected the $k_1^-$ term. Using Equations (11) and (6) the growth rate can then be written as,

$$R_G = k_2^+[\mathrm{C_{ads}}] = 2k_1^+ K_0 \frac{[\mathrm{C_2H_4}]}{[\mathrm{H_2}]} \quad (18)$$

From this result we can note that the growth rate is similar to the surface diffusion-limited regime except it is now only depending on $k_1^+$ and $K_0$, as expected since $k_1^+$ is the dissociation term.

**Mass diffusion-limited regime.**

Finally, we consider the mass diffusion regime, which also occurs at a high ethylene partial pressure. However, this regime is characterized by very low flow velocity around the catalyst so that the mass diffusion of ethylene to the catalyst becomes the rate-limiting step. In this case, the consumption of ethylene exceeds its supply by diffusion. On increasing partial pressure of ethylene, the diffusion constant decreases which can finally result in an undersupply of carbon at the catalyst.[19] This regime is characterized by a low activation energy.[17]

## Results and Discussion

With the different proposed regimes, combined with the rate equations for CNT growth, we can investigate the rate-limiting steps as proposed in Fig 1. A wide range of experimental conditions is considered in order to study and optimize the carbon nanotube growth. We compare the quality and morphology of the resulting growth as a function of temperature and partial pressure of the different process gases, linking the results to the growth regimes described in the previous section. The fast *in situ* growth of our laser-assisted CVD method allows us to rapidly explore a wide range of different experimental conditions and to determine the activation energies for the different rate-limiting mechanisms.

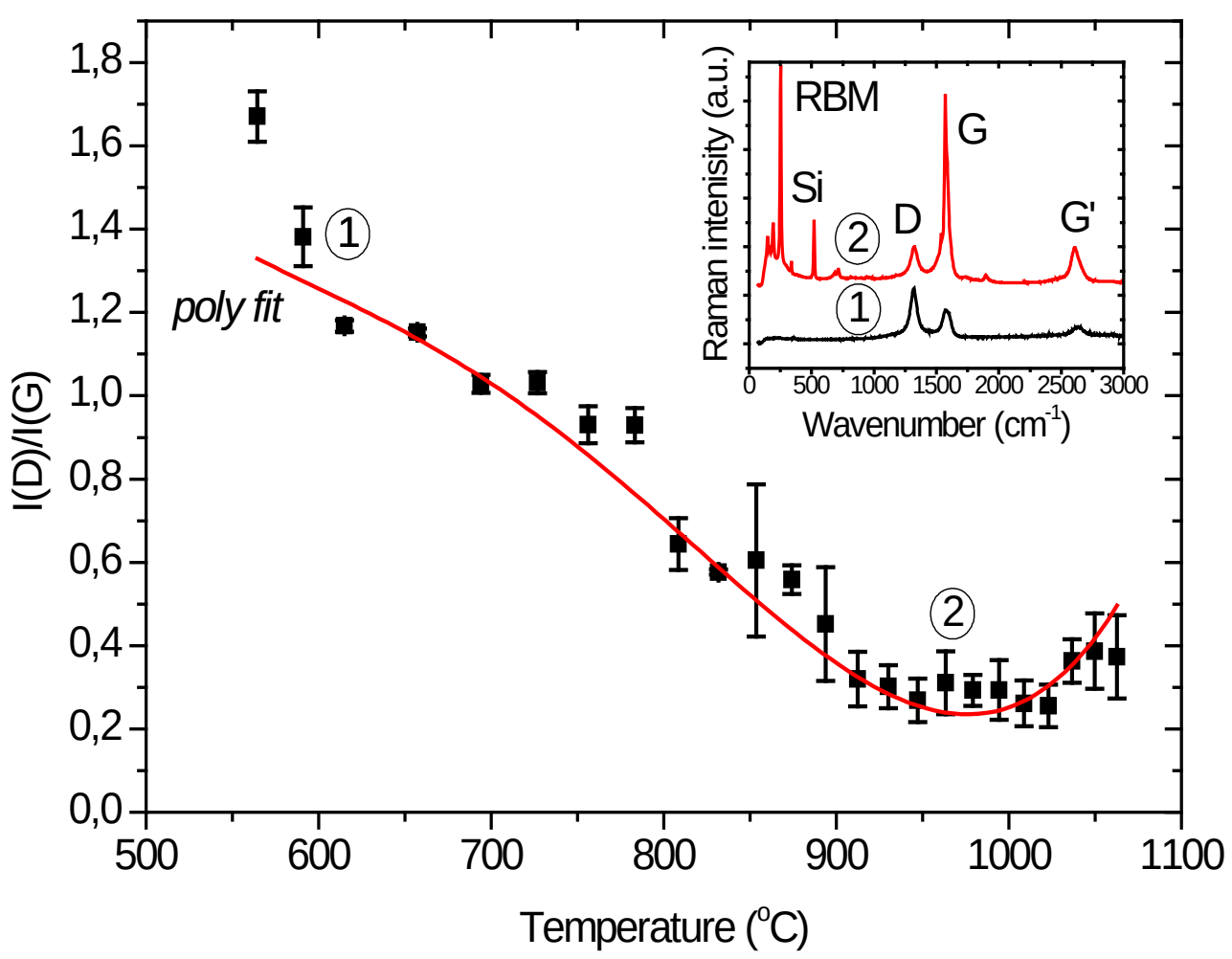

Fig 4. Estimated process temperature versus I(D)/I(G) ratio. The figure inset shows typical Raman intensity spectra for two different experiments specified in the figure. Indicated in the figure are characteristic peaks for the radial breathing modes (RBM), Silicon (Si), the defect-peak (D), the graphite-peak (G) and the second order graphite peak (G').

To investigate the influence of temperature, we rely on thermal modelling of the process, described elsewhere[13] (see Methods for more details). For ideal gases, the rates of reaction are dependent on the temperature $T$, the universal gas constant $R$ and the activation energy $E_a$ following Arrhenius law,[25]

$$k(T) = ve^{-E_A/RT} \qquad (19)$$

where $v$ is the frequency factor. Similarly, the equilibrium constant $K$ is dependent on the temperature and gas constant through the following relation,

$$K(T) = e^{-\Delta H/RT} \qquad (20)$$

In general, the quality of the nanotubes grown with a CVD process with ethylene and iron increases until about 900°C.[26,27] A similar trend is found in Fig 4, where the ratio of D and G intensity is plotted against estimated process temperature. The data is fitted with a polynomial fit and shows a minimum around 950°C. The figure inset shows two typical Raman spectroscopy graphs, with graph 1 showing low quality mostly multi-wall CNTs while graph 2 clearly shows the presence of single-walled high quality CNTs.

The decrease in the quality after a certain temperature can be attributed to an increase in the pyrolysis of ethylene. At high temperature the non-catalytic decomposition of ethylene increases, usually in the form of amorphous carbon deposition. This trend can also be a consequence of the increased reaction rate in the surface diffusion-limited regime. Higher temperatures can also lead to a more efficient dissociation and thickening of the nanotube walls, introducing defects and decreasing overall quality of the CNTs.[28] Although the growth rate and quality generally are higher for higher temperatures, at very high temperatures the resulting growth morphology consists of predominantly un-aligned CNTs forming a thin spaghetti-like nano-carpet. Most probably a few high quality single-wall nanotubes contribute to the low I(D)/I(G) ratio and the growth terminates quickly, hampering the formation of forests. In the growth rate study we therefore assume the structure of the CNTs to be more or less identical *i.e.* aligned multi-wall CNTs.

**Growth Rate and Activation Energy**

The CNT growth rate is proportional to the reaction rate $k$ and the concentration or partial pressure of adsorbed carbon, according to Equation (11). Using Equation (19) the activation energy of the CNT growth process, $E_a$, is found. Using the natural logarithm of the growth rate plotted against the inverse of the temperature, Arrhenius law is used to find the activation energy using

$$R_G \sim e^{-E_a/RT} \qquad (21)$$

In this case, $R_G$ is the CNT growth rate and $E_a$, is the apparent activation energy, since it includes all elementary steps included in the carbon nanotube growth. Although the precise underlying mechanism is still under debate, different groups have determined the rate-limiting step and activation energy for carbon nanotube growth.[16,17,19–22,29–47]

Baker *et al.* argued that the rate-limiting step must be carbon diffusion in bulk iron resulting from the resemblance of the activation energy of bulk diffusion with their calculated activation energy.[29] Other groups found similar results relating the activation energy to bulk carbon diffusion[31,34,39] as the activation energy was between 1.3 and 2 eV corresponding to the bulk diffusion energy of fcc-iron of 1.53 – 1.57 eV.[48,49] On the other hand, diffusion of carbon in bulk bcc-iron has an activation energy of about 0.8 eV.[48,49] In contrast, other research has shown that the surface diffusion of carbon has a much lower activation energy of about 0.2 – 0.3 eV and is thus energetically favorable.[20,30]

The previous results must be a combination of both dissociation as well as diffusion, as was also proposed by Bronikowski *et al.*[37] The low values of 0.2 – 0.3 eV were also found in fluidized bed reactors.[41,43] Nessim *et al.*[42] found a very low activation energy of 0.1 eV if the process was carried out without pre-heating of the process gases while with pre-heating an activation energy of 0.9 eV was reported.

Liu *et al.*[21] and Pirard *et al.*[22] argued that the mechanism must be surface reaction- or dissociation- limited and reported activation energies of 1.65 eV and 1.25 – 1.4 eV respectively. In comparison, acetylene and ethylene dissociate on the iron

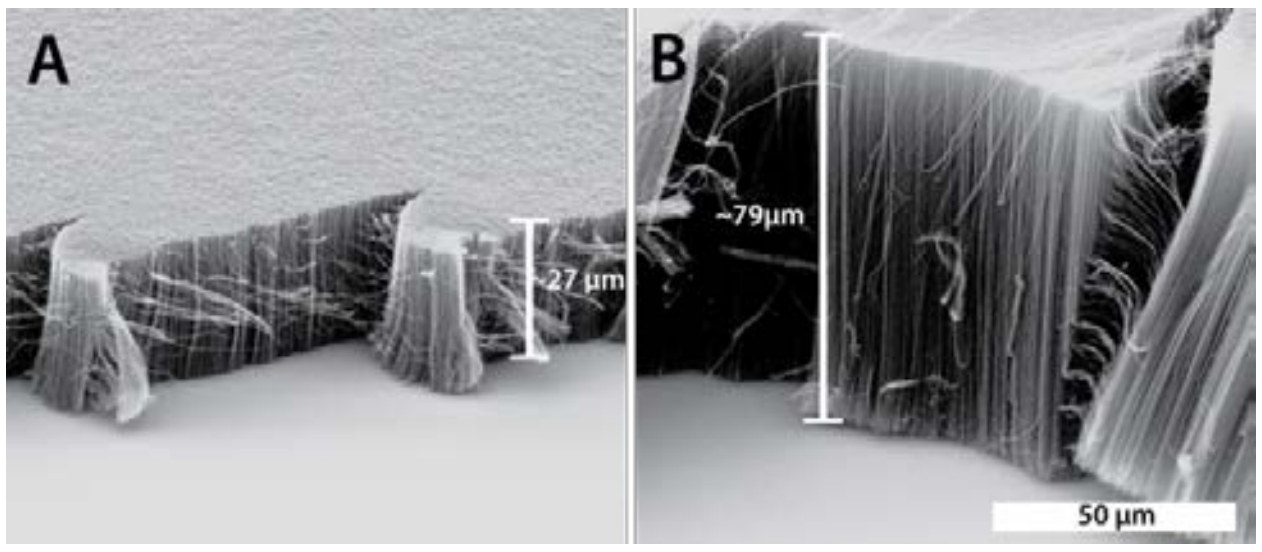

Fig 5. SEM pictures showing vertically aligned nanotube forests grown under similar conditions with only partial pressure of ethylene changed. A. Flow of ethylene 20 sccm (partial pressure = 0.07 bar). B. Flow of ethylene 250 sccm (partial pressure = 0.5 bar).

surface via $CH_x$ radicals with an activation energy of about 0.6 – 1.4 eV.[50]

Without experimental estimation of the activation energy, mass diffusion was proposed as the limited process by Zhu *et al.*[23] and Louchev *et al.*[35]

These results illustrate the diversity between the attributed underlying mechanisms that determine the apparent activation energy. It is fair to say that the growth process is not fully understood and the variation of the activation energy between 0.1 and 2 eV reflect the strong influence that different process parameters have on the growth mechanism. The direct *in situ* laser-assisted growth process proposed in this paper offers a quick and versatile method to investigate activation energy for multiple growth regimes.

### Influence of Ethylene

Using a constant flow of argon and hydrogen, the influence of the flow of ethylene on the growth has been investigated. The partial pressure of ethylene was increased from 0.04 bar to 0.7 bar. This was achieved by flowing argon and hydrogen at a constant flow of 200 and 50 sccm respectively and changing the flow rate of ethylene from 10 to 500 sccm. From Equations (10) and (11), we expect that a higher ethylene partial pressure leads to higher growth rates. To confirm this, we have estimated the length of the CNTs using SEM. This is shown in Fig 5. It is clear that the CNT length is lower for a partial pressure of 0.07 bar (ethylene flow of 20 sccm) than for 0.5 bar (ethylene flow of 250 sccm).

By comparing Raman spectroscopy intensity signals we found a clear trend for increasing ethylene partial pressure. Higher pressures, thus higher growth rates, results in lower CNT quality. This could be attributed to defect formation at high growth rates.[28] An additional reason could be the transition to a different regime with a higher density of defects due to the oversupply of carbon and poisoning of the catalyst. In Fig 6, the quality of nanotubes is shown as a function of the ethylene partial pressure for two different temperatures and argon flow rates. In this figure, a distinction is made between *surface diffusion-limited* growth and *dissociation-limited* growth. The transition between both regimes is positioned around a partial pressure of ethylene of 0.2 bar. For the lower temperature graph (in red), the decrease in quality starts at a slightly lower ethylene partial pressure, resulting from the lower total flow rate. The overall quality is also lower, in accordance with Fig 4.

We calculated activation energies for both regimes using Arrhenius law, see Fig 7. We found activation energies for the surface diffusion-limited regime and for the dissociation-limited regime of 0.8 and 0.6 eV, respectively. The dissociation-limited regime has the lowest activation energy of about 0.6 eV with a growth rate that is generally higher, in accordance with Equation (18). This seems realistic, as the dissociation step is always present and the surface diffusion can vary as a result of the total coverage of carbon on the catalyst surface. High temperatures are excluded from the analysis as they show a sudden decrease in growth rate, a result of the non-aligned growth as discussed before.

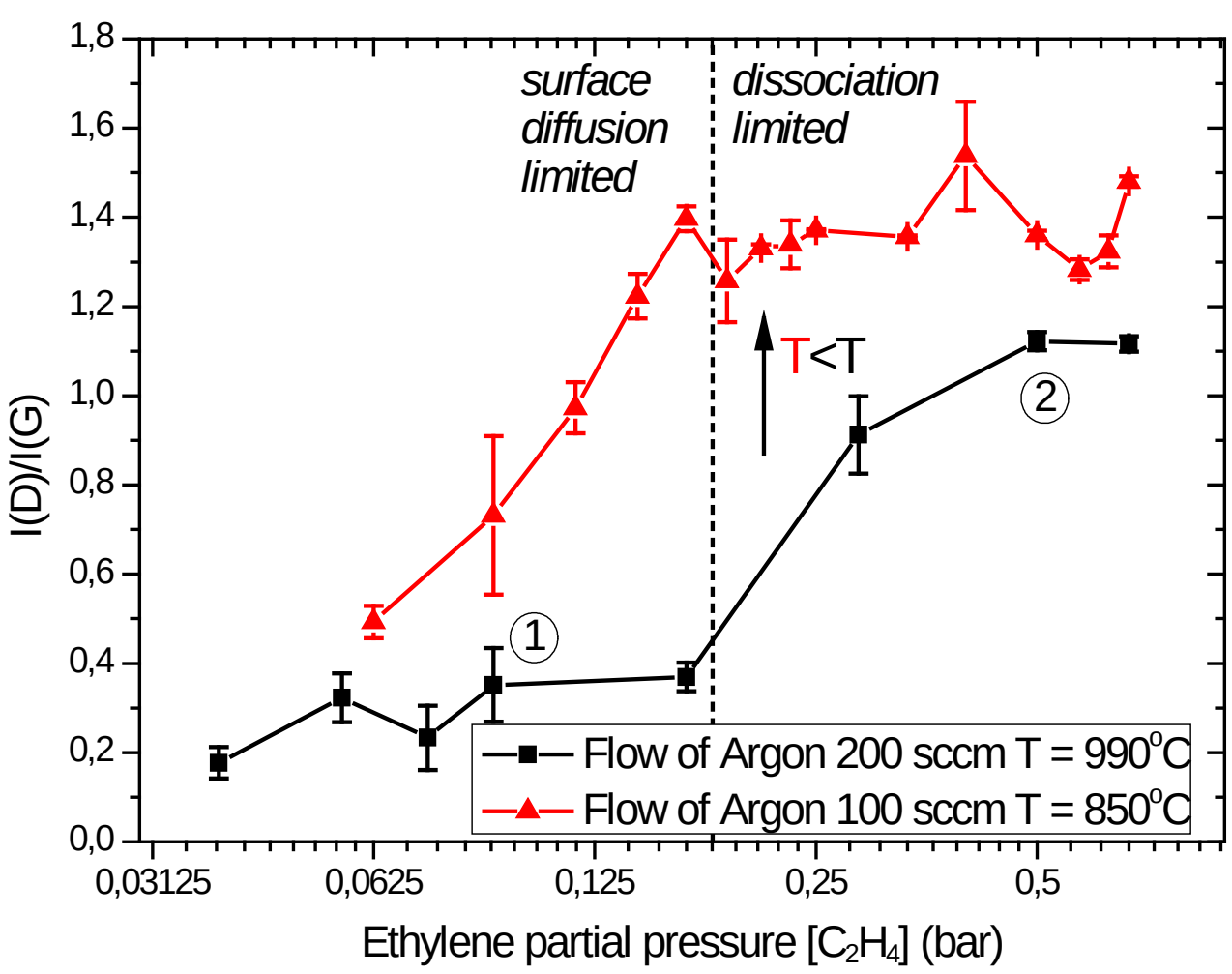

Fig 6. Quality of grown CNTs plotted versus ethylene partial pressure for two growth temperatures. Different growth regimes are indicated in the figure.

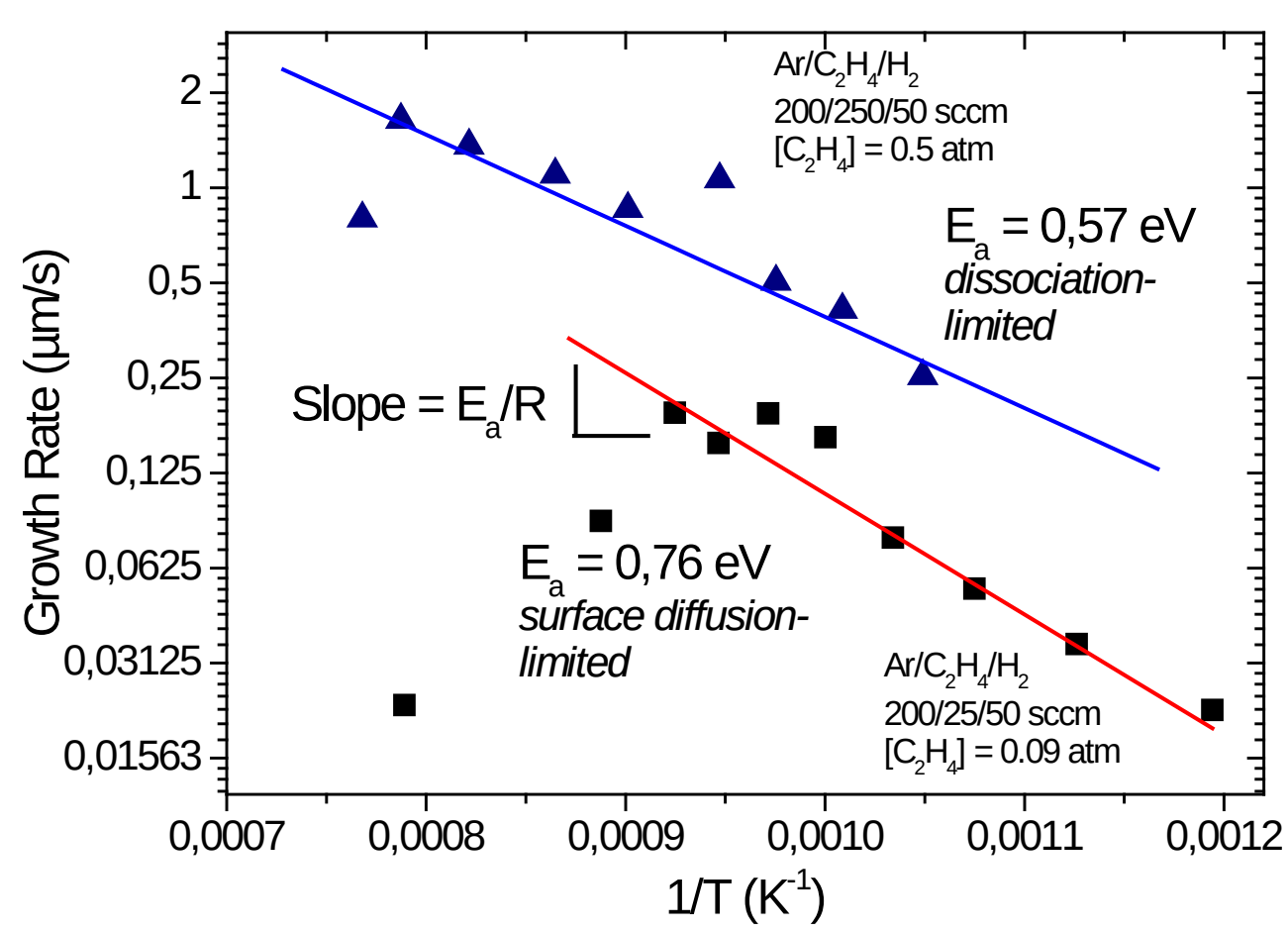

Fig 7. Arrhenius plot of the surface diffusion-limited and the dissociation-limited regimes. The CNT growth rate is shown versus the inverse of the growth temperature. Indicated is the partial pressure of ethylene for both graphs as well as the experimental flow conditions.

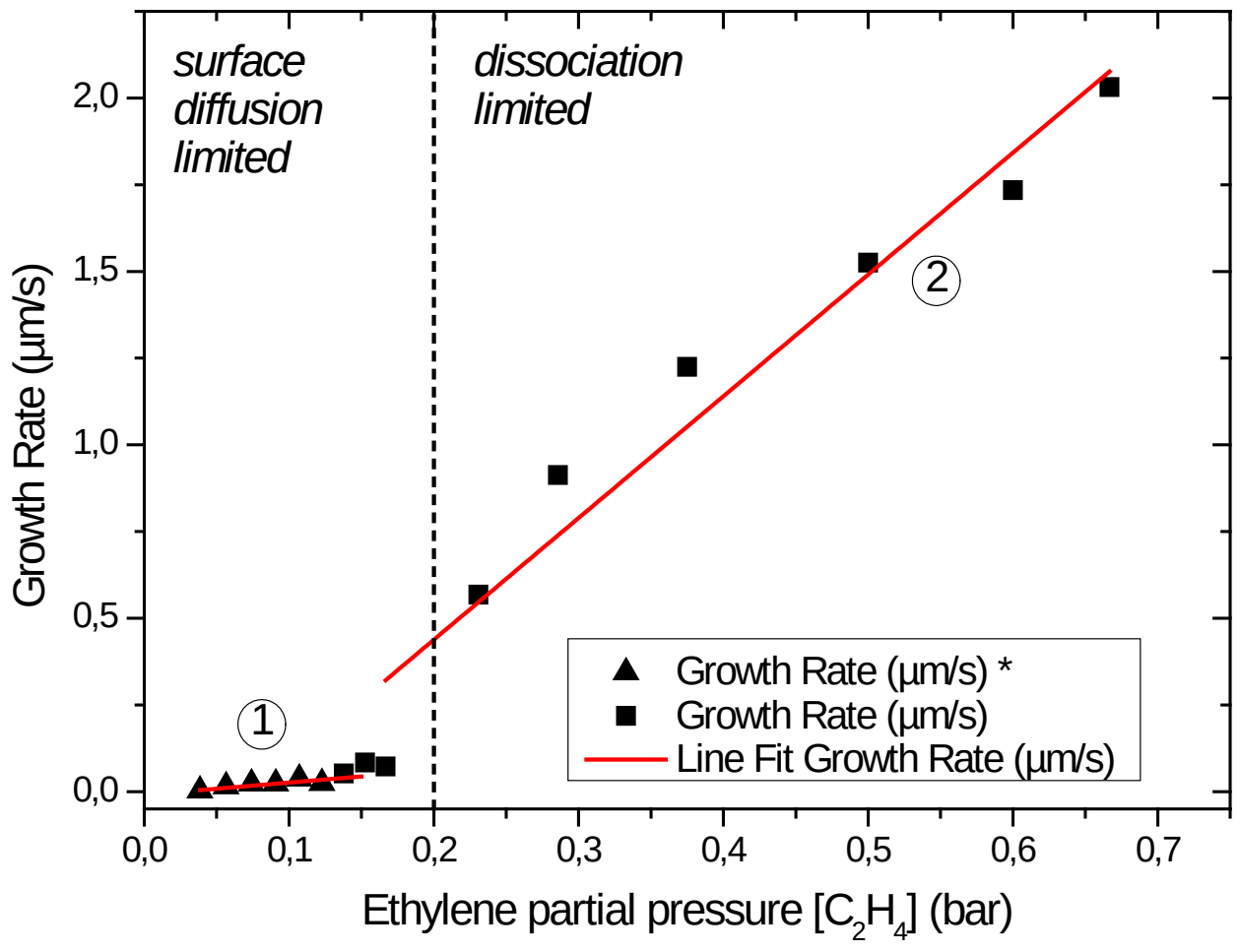

Fig 8. Growth rate versus ethylene partial pressure. In both regimes a linear relation was found and the reaction order is therefore assumed to be 1. *Growth rate is calculated using a less significant drop in reflection, see Methods for more details. Experiments 1 and 2 correspond to experiments 1 and 2 in Fig 6.

Our calculated activation energies are lower compared with other reported activation energies for surface diffusion- and dissociation-limited processes. However, using plasma-enhanced CVD, Hofmann *et al.*[20] also reported a very low activation energy, which was attributed to surface diffusion only; the plasma involved accounted for the decomposition of the ethylene. We cannot exclude such effects from our experiments. Another possibility is that in our process the catalyst is quickly transferred into iron-carbide, $Fe_3C$, where the diffusion of carbon is calculated to be 0.38 eV by Liu *et al.*[21] Sharma *et al.*[51] have proposed that the growth of CNTs takes place on cementite that is formed during the reduction process of iron-oxide. A more detailed study of the catalyst chemical composition is necessary to confirm or disprove these assumptions. Finally, the results exhibit uncertainties due to errors introduced by fitting with this amount of data points[45] and the average growth rate used in the calculations could also account for some deviations.

The reaction rate order of ethylene is investigated by measuring growth rate as a function of ethylene partial pressure. The results are illustrated in Fig 8. For both regimes we found a linear relation and as such we assume the reaction order to be 1, in correspondence with other reported results[19,39,46] and Equation (18). The surface diffusion-limited growth has a much lower linear relation to the ethylene partial pressure as the dissociation-limited growth, which corresponds to both a lower reaction rate constant *k* and the weaker dependence on ethylene partial pressure for diffusion-limited growth.[21]

**Influence of Argon**

The rate equations do not include a term for the partial pressure of argon. In fact, the flow of argon and the partial pressure do not influence the ratio of the partial pressures of ethylene and hydrogen, as they will both be changed equally by any change of partial pressure of argon and as such Equation (15) will not change. However, the carrier gas argon still influences the growth of the carbon nanotubes. The dilution of the process gases by argon can indeed change the rate limiting process steps. In effect, the argon flow rate determines the flow speed through the chamber. At the same time, a high argon flow results in a low concentration of both ethylene and hydrogen which may result in an undersupply of those gases at the catalyst site. In that case, the argon molecules will shield the surface sites, decreasing the mean free path of ethylene molecules, effectively preventing them to reach the catalytic surface. On the other hand, with no argon flow, the flow velocity is low which could result in a local undersupply of carbon, when consumption is higher than supply. For both a low and a high argon partial pressure we have analysed the resulting growth. This can be seen in Fig 9. The SEM pictures show the growth for similar experimental conditions where only the argon flow is changed and demonstrate the similarity of the growth rate of both cases. The insets in the figure show the morphology of the aligned CNT forest which indicates that although the growth rate might be comparable; the morphology of the CNTs is not.

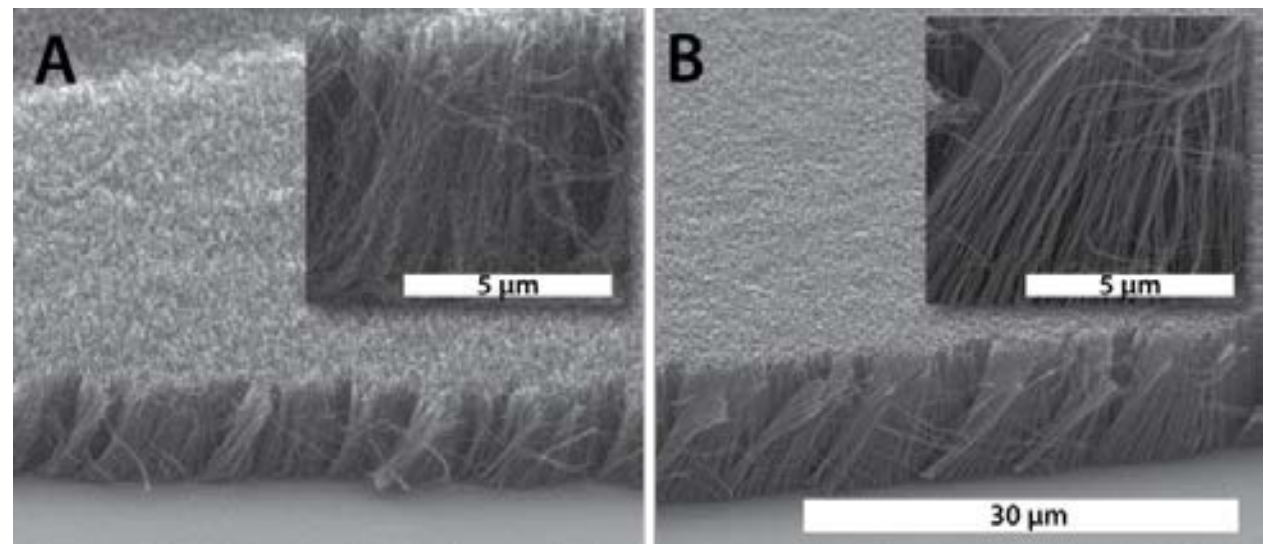

Fig 9. SEM pictures showing influence of argon flow and partial pressure on growth rate. A. Argon flow 10 sccm (partial pressure = 0.14 bar). B. Argon flow 500 sccm (partial pressure = 0.9 bar). Both insets show a detailed zoom of the morphology of the aligned nanotube forest.

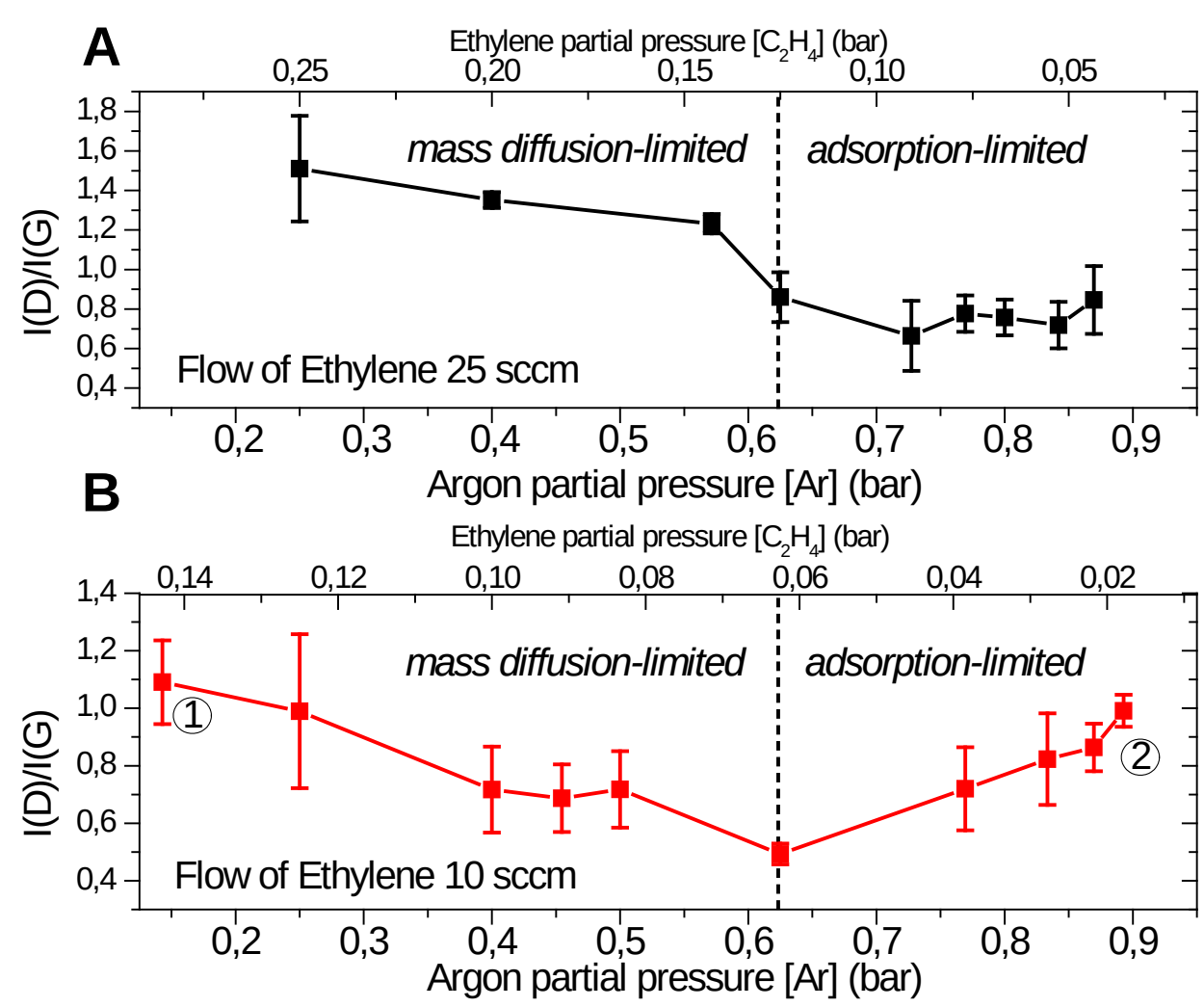

Fig 10. Quality of nanotube growth versus argon partial pressure. Indicated are also the ethylene partial pressure and the separation between both regimes. Experiments 1 and 2 correspond to the SEM pictures of Fig 9(a) and (b), respectively.

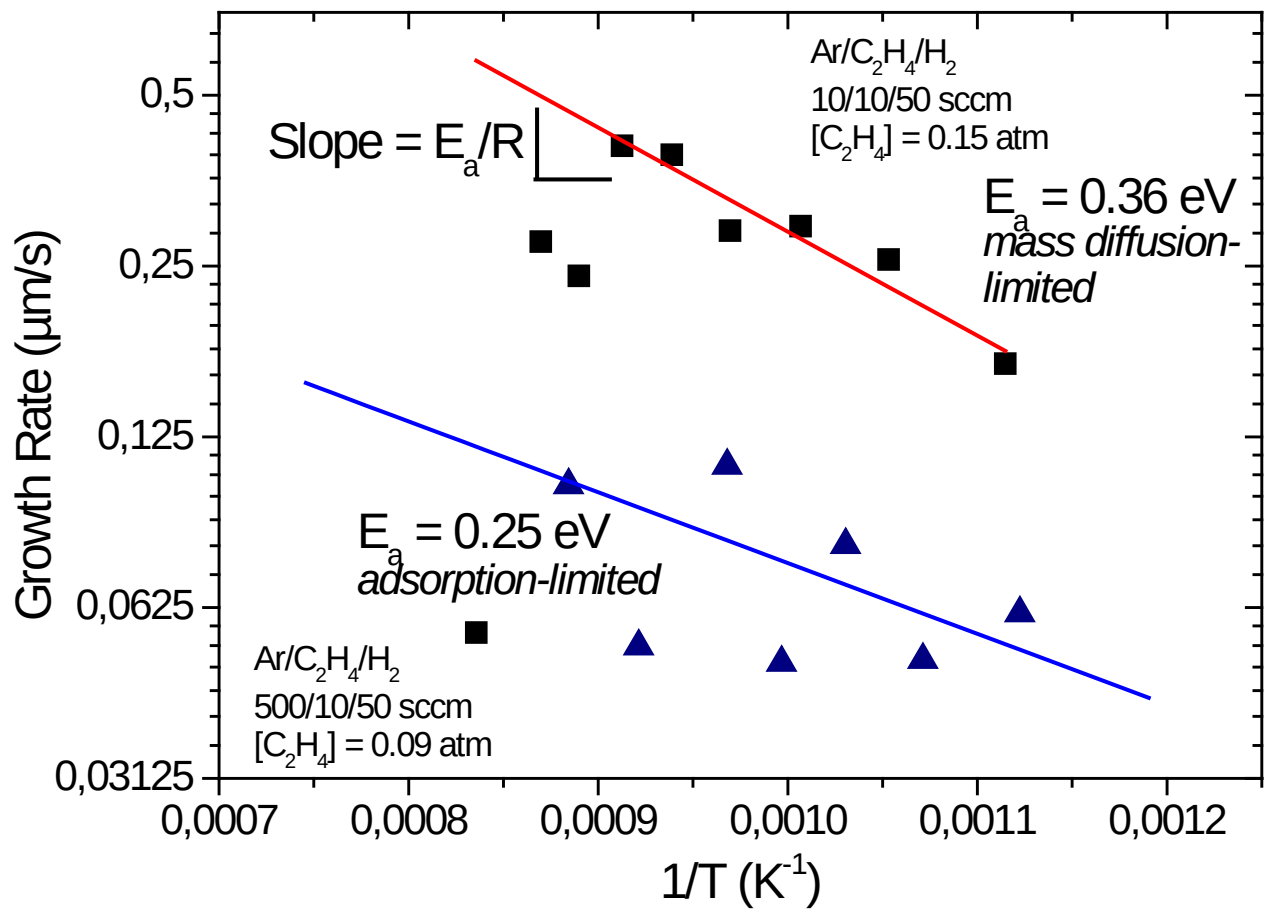

Fig 11. Arrhenius plot of the mass diffusion-limited and the adsorption-limited regime. The growth rate is potted against the inverse of the calculated growth temperature. The partial pressure of ethylene are indicated for both graphs as well as the experimental flow conditions.

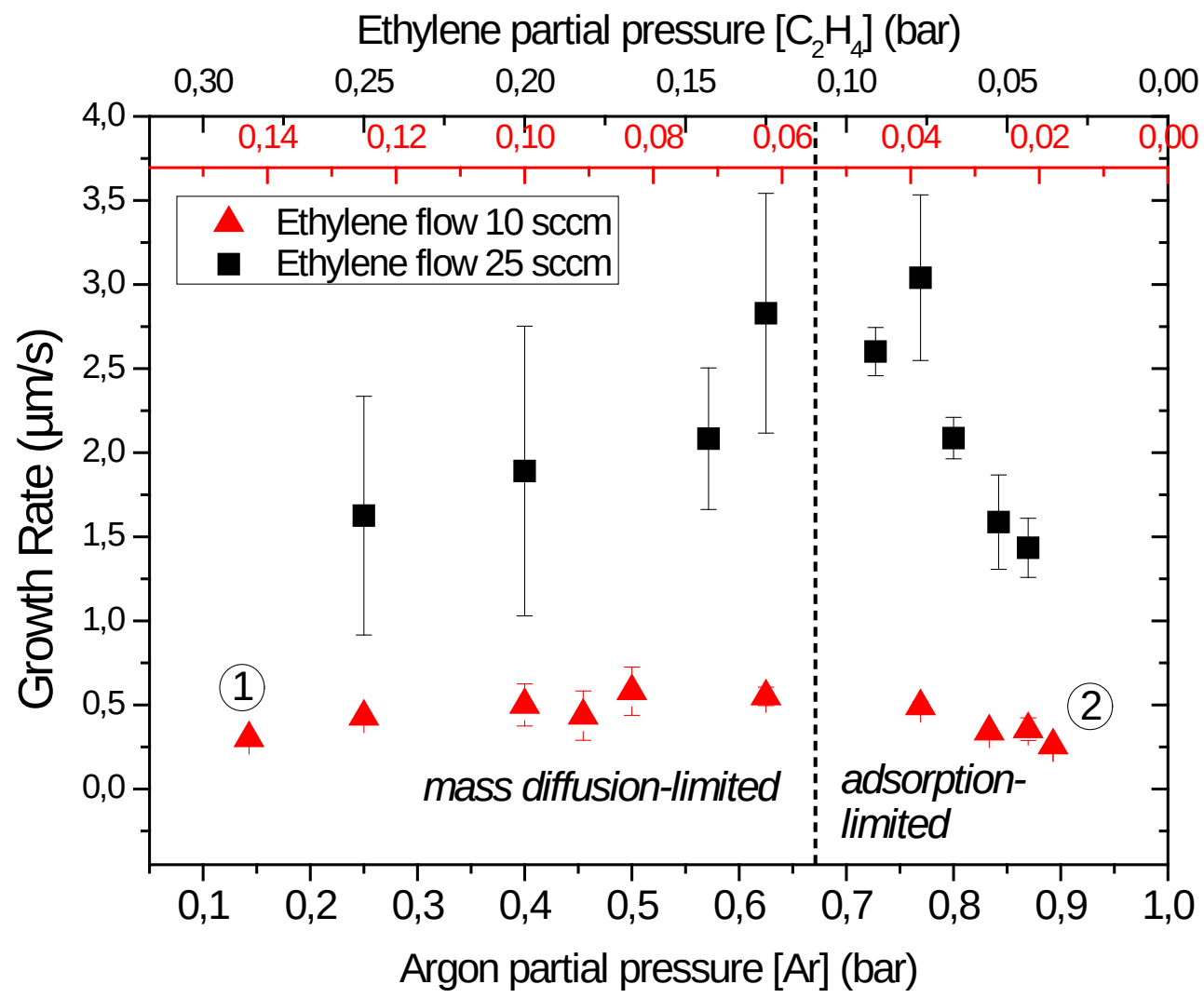

Fig 12. Growth rate as a function of argon and ethylene partial pressure for two different ethylene gas flows, 10 and 25 sccm, respectively. From the graph it is clear that argon has negligible influence on the growth rate. The growth rate for the ethylene flow of 25 sccm is larger than that of 10 sccm and larger errors are present in the mass-diffusion limited regime. Indicated in the figure is the transition between the mass diffusion-limited and adsorption-limited regime. Experiments 1 and 2 correspond to experiments 1 and 2 in Fig 10 as well as Fig 9(a) and (b), respectively.

In Fig 10 the quality of the CNTs is assessed as a function of argon partial pressure. The other two gases, ethylene and hydrogen, are kept constant. The quality of CNTs is assessed for two different ethylene flows, 10 sccm, Fig 10(a) and 25 sccm, Fig 10(b), respectively while the hydrogen flow was 50 sccm throughout.

There is a clear trend visible in both graphs. For low argon partial pressures, the quality of the nanotubes is low. This is attributed to the mass diffusion-limited regime where the local consumption of ethylene by the catalyst is higher than the supply of ethylene by diffusion as a result of the very slow flow rate of gases around the hot-spot. In this case that means $k_0^+ < k_1^+$. With insufficient carbon in the catalyst, the growth is not stable which results in low quality CNTs. An optimum value of the quality is found around 0.65 – 0.75 bar argon partial pressure where we assume the change between growth regimes is approximately located.

In contrast, an increase in argon flow to high partial pressure also leads to a reduction in the quality. We suspect this regime is dominated by the adsorption-limited regime. With a high argon flow, the surface coverage of the ethylene is low and the not enough carbon atoms are absorbed to ensure a stable growth. This results in both low quality nanotubes as well as lower growth rates. This effect seems to be larger for the ethylene flow of 10 sccm which is plausible considering the amount of ethylene molecules around the catalyst is much lower in that case.

To qualitatively assess both regimes, the activation energy for experimental conditions 1 and 2 from Fig 10(b) is calculated. The result is shown in Fig 11 where both regimes are fitted with an Arrhenius plot. The resulting calculated activation energies are very low. This might be the result of the large errors in the growth rate fit; the adsorption-limited regime in particular shows large errors around the fit. We assume that a high sensitivity on process conditions is the cause for the large errors. In the adsorption-limited regime, the argon partial pressure is very high, preventing ethylene molecules from reaching the catalyst surface; small deviations in process conditions might have a large impact on the growth rate.

The mass diffusion-limited regime is expected to have a low dependence on temperature[19] and thus a low activation energy. Because the driving force is diffusion by concentration difference it is an unstable regime and also sensitive to small deviations in experimental conditions.

The reaction rate order of argon is investigated by plotting the growth rate as a function of argon partial pressure (Fig 12). In the figure the proposed growth regimes from Fig 10 are indicated. From the figure it is clear that the reaction rate order is not zero but the influence on the growth rate is minimal in the case of an ethylene flow of 10 sccm. For the higher ethylene flow, an increase in argon partial pressure leads to a slight increase in growth rate. This can be explained by the increased flow rate, ultimately overcoming the diffusion limited growth by forced flow of new ethylene supply. At high argon partial pressure and a high flow rates, a visible reduction of the growth rate predicts a reaction order of -1, which can be explained by the increased coverage of the argon molecules on the catalyst, preventing the ethylene from adsorbing and dissociating on the catalyst.

**Influence of hydrogen**

Hydrogen has previously been found to be a key player in the CNT growth process. For instance, it plays a significant role in the reduction process of the iron-oxide catalyst layer and is an excellent thermal conductor and as such has been used as a carrier gas, optimizing growth kinetics. Hydrogen has also been suggested to decompose inactive metal carbides[52,53] and plays a significant role in the removal of amorphous carbon from the catalyst surface, preventing catalyst poisoning and increasing the process lifetime.[54,55] In contrast, hydrogen has also been reported to suppress the CNT growth by the formation of methane from carbon.[52,53]

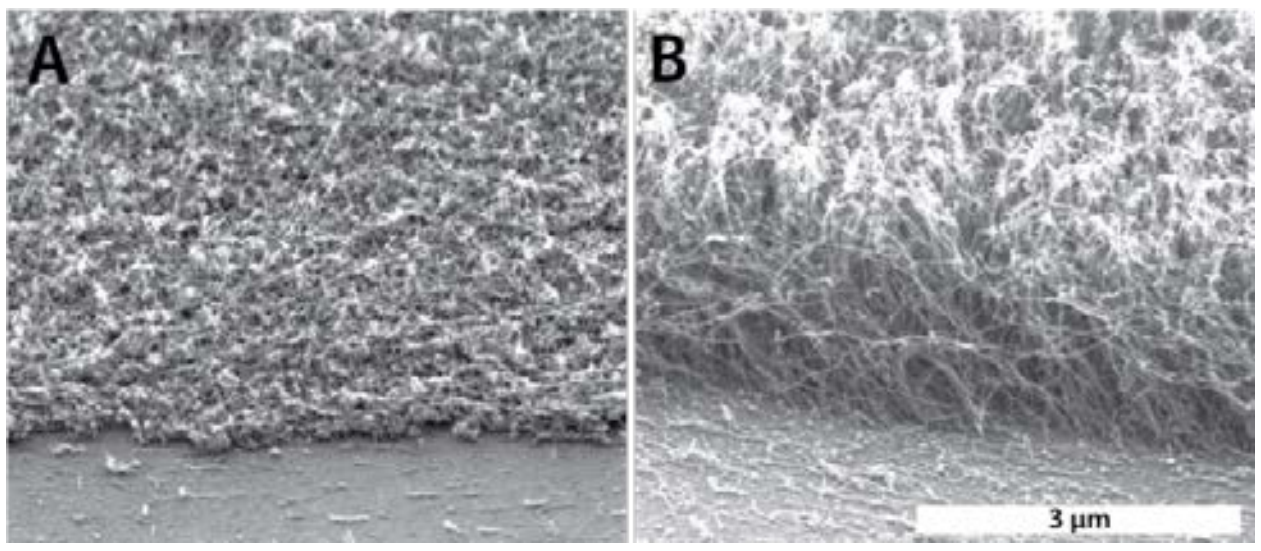

Fig 13. SEM pictures of the influence of hydrogen on resulting CNT growth. A. Hydrogen flow 400 sccm (0.64 bar). B. Hydrogen flow 25 sccm (0.1 bar).

The hydrogen incorporated in the ethylene gas can also be used in the reduction process at high temperature, therefore we have investigated the influence of hydrogen on CNT morphology and quality by varying the flow between 0 and 200 sccm (partial pressure 0 – 0.47 bar) while keeping the argon and ethylene flow at 200 and 25 sccm respectively. From the growth rate equations, Equations (10) and (11) we suspect the hydrogen partial pressure negatively influences the growth rate. In Fig 13, two SEM pictures are presented with similar growth conditions except for the hydrogen partial pressure. A high hydrogen partial pressure, Fig 13(a), results in a very thin layer of nanotubes while a low hydrogen partial pressure, Fig 13(b), results in a forest of longer CNTs.

Since hydrogen is involved in multiple stages of the growth process, the quality of the CNTs is also expected to be influenced by the hydrogen partial pressure. In Fig 14, we have plotted the quality of CNTs as a function of hydrogen partial pressure and ethylene partial pressure for two durations of laser exposure, 150 and 20 seconds respectively.

A small shift in the minimum I(D)/I(G) ratio can be noticed between the two graphs towards the lower hydrogen partial pressure. Although the overall quality of the 150 second experiments is generally better, the best quality is reached at a higher hydrogen partial pressure than in the 20 second case. As an explanation, we look back at Equation (10) and (11). The exponential term goes to zero for sufficiently large *t*. However, the hydrogen partial pressure also plays a role in this term. The difference between both experimental times is that the extra time in the longer experiment, will ensure that the growth rate is stabilized, *i.e.* an equilibrium reached, even for low hydrogen partial pressures, and will result in high growth rates due to the inverse relation of hydrogen. High growth rates can introduce more defects. On one hand, a low hydrogen partial pressure for the 20 second case might still be outside equilibrium, resulting in a lower growth rate and more defect free growth. On the other hand, a high hydrogen partial pressure will result in lower growth rates and generally higher quality CNTs. However, the resulting ethylene partial pressure also drops, *i.e.* the coverage of the catalyst surface by ethylene molecules becomes very low which could result in a transfer to the adsorption-limited regime, similar to the case of high argon partial pressure.

Since hydrogen pre-dominantly determines the start and nucleation of the growth process, due to its influence in reduction of the oxidized catalyst and the exponential term in Equation (10), we investigate the nucleation rate of the process as a function of hydrogen partial pressure. The temperature set-point was set to ~950°C. The experiment already starts (and all three gases are present) during the ramping to this set-point to investigate nucleation characteristics. The nucleation rate is defined as the inverse of the nucleation time $t_n$ and can be extracted from Equation (10) by introducing a nucleation constant $C_N$,

$$N_R = C_N \frac{[C_2H_4]}{[H_2]}\left(1 - e^{-k_2^+[H_2]t}\right) \tag{22}$$

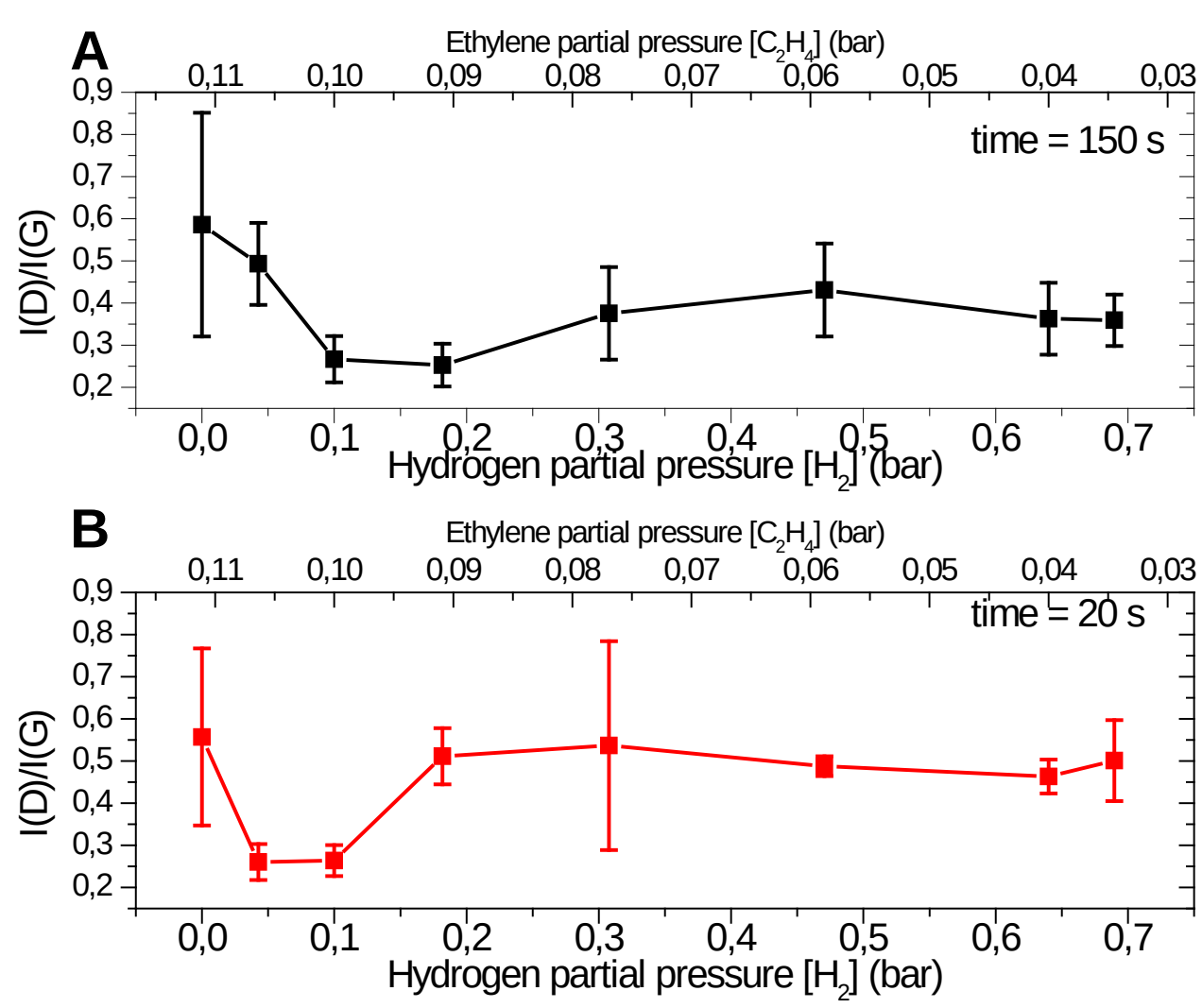

Fig 14. Quality of CNTs as a function of hydrogen and ethylene partial pressure for two experimental growth times: (a) 150 seconds and (b) 20 seconds.

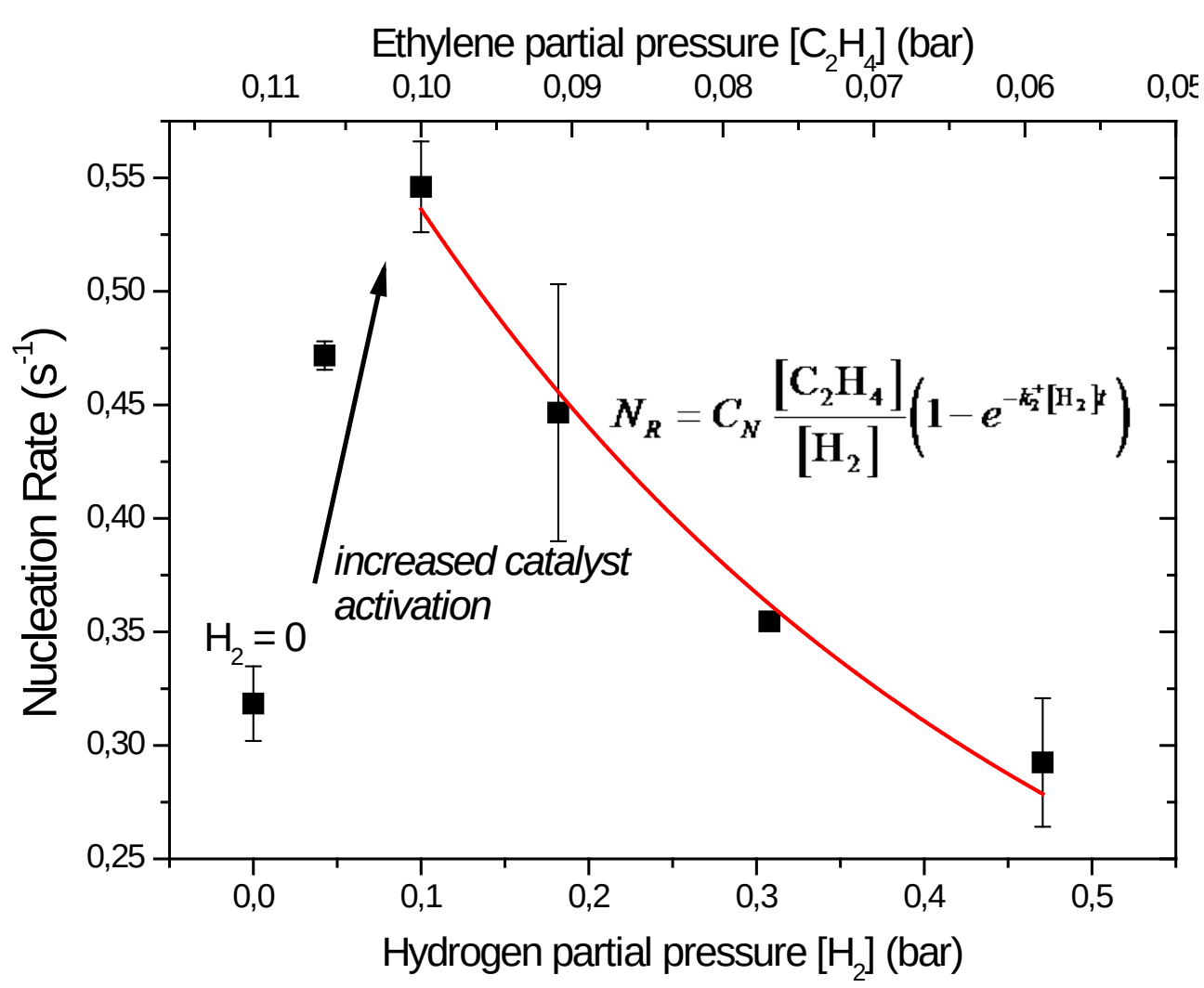

Fig 15. Nucleation rate of carbon nanotube growth as a function of hydrogen and ethylene partial pressures. From the rate equations an exponential relation is expected as depicted by the red fitted line. However, no hydrogen would incorrectly result in an infinite nucleation rate. Hydrogen serves in reduction of oxides and is necessary for the catalyst activation. Therefore, first an increase in nucleation rate is noticeable followed by a decline that follows the trend of the exponential decay.

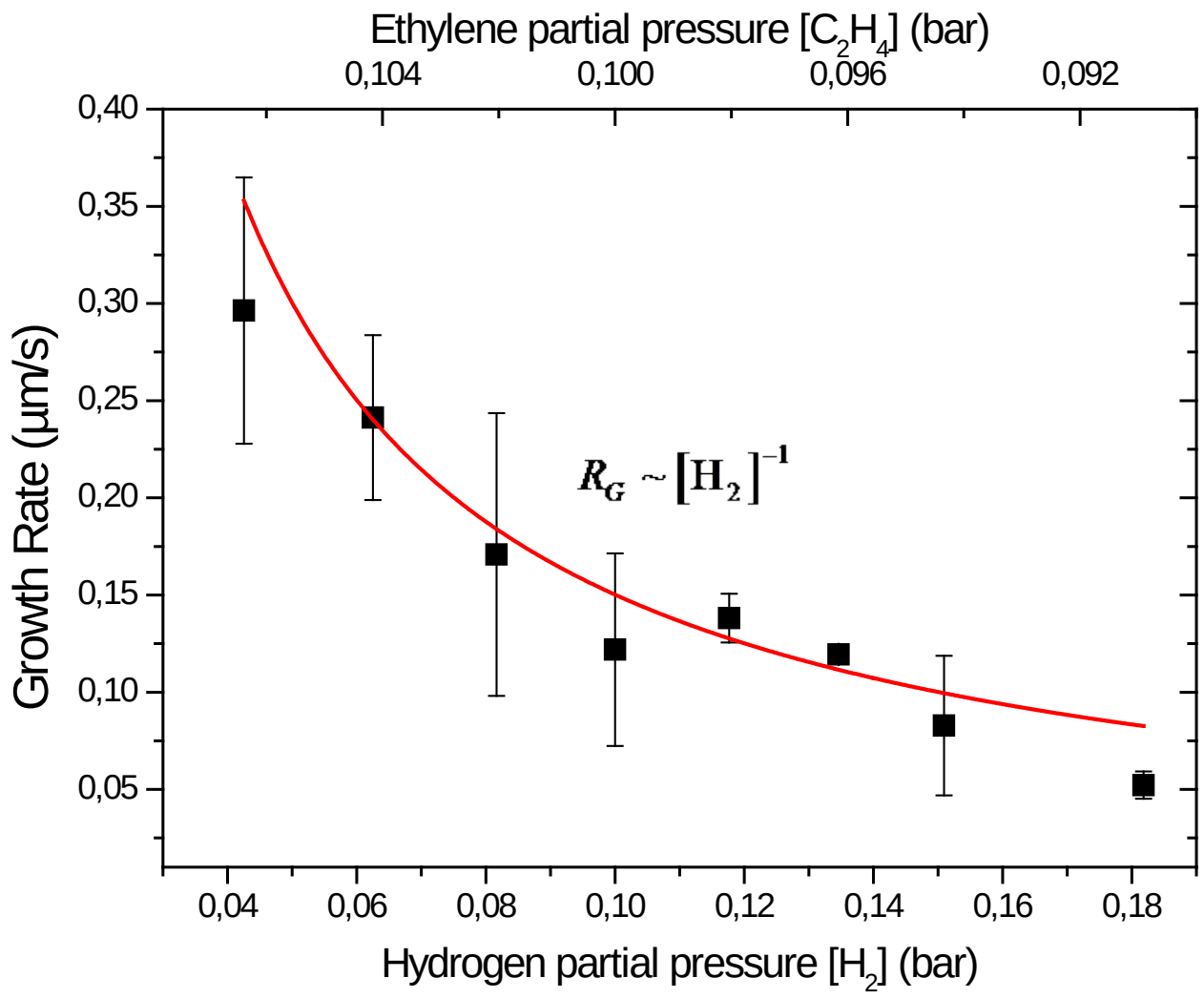

Fig 16. Carbon nanotube growth rate as a function of hydrogen partial pressure. Also indicated is the resulting ethylene partial pressure. The growth rate shows an inverse relation to the hydrogen which is fitted by the red line.

The nucleation rate is plotted against the hydrogen partial pressure in Fig 15. Two parts can be distinguished in this figure. The first (left) part shows an increase in nucleation rate while the second (right) part shows an exponential decay. The decline can be explained by fitting Equation (22). However, in this equation no hydrogen would result in an infinite nucleation rate. Since hydrogen also plays a role with the reduction of the catalyst, the first part of the figure can be explained by an increase in catalyst activity. Without hydrogen, the ethylene has to provide the hydrogen, which can only decompose thermally at high temperatures. So the catalyst is reduced faster with more hydrogen, until the point at which more hydrogen would start suppressing the nucleation rate. A similar trend for the influence of hydrogen concentration was found by Pérez-Cabero *et al.* for acetylene on iron catalyst.[56]
The reaction rate order of hydrogen is estimated in Fig 16 where the hydrogen flow is varied between 10 and 50 sccm (partial pressure 0.04 – 0.18 bar). The ethylene gas is added after the temperature set-point is reached to exclude the influence of start-up and nucleation. From Equation (10) we suspect an inverse relation between growth rate and hydrogen partial pressure, *i.e.* growth rate $\propto [H_2]^{-1}$. In the figure the growth rate is fitted with this relation, which shows that the reaction rate order of -1 proposed in the rate equations seems probable. Note that for these experiments the growth temperature was too low (~850°C) to result in CNT growth without any hydrogen.

## Conclusion

In this paper we investigated the growth kinetics of laser-assisted carbon nanotube growth. The localized nature of the laser-assisted process allows for a fast *in situ* determination of growth kinetics with high throughput with respect to conventional CVD. As a result, we were able to investigate the influence of the process gases over a wide range of partial pressures and propose several growth regimes. Each of the growth regimes has a unique rate-limiting step which is determined by the governing rate equations. The quality and morphology of the CNTs is assessed as a function of temperature and partial pressure for ethylene, argon and hydrogen. The temperature shows a maximum CNT quality around 950°C. The growth rate is calculated using the dynamically changing reflected intensity from the laser beam. From the growth rate we could determine reaction rates and activation energies. The activation energies found are in the range between 0.3 – 0.8 eV. The results are linked to growth regimes, depending on the partial pressure and flow rate of the different process gases.
A variation in the partial pressure of ethylene shifts the growth regime from surface diffusion-limited to dissociation-limited growth. The quality decreases with increasing ethylene partial pressure while the growth rate increases. This is confirmed by SEM and Raman analysis, which indicates the largest forest to be low quality multi-walled CNTs grown in the dissociation-limited regime. The reaction rate order of ethylene is found to be 1, in accordance with derived rate equations.
The partial pressure of argon in itself does not change the growth rate but in effect changes the flow speed of the process gases over the substrate. With a very low argon partial pressure the mass-diffusion of ethylene to the growth site is the rate-limiting step. The quality of the growth decreases with lower partial pressure. In contrast, a high argon partial pressure can affect the surface coverage of the catalyst. That is, the argon molecules shield the surface sites, preventing the ethylene molecules to reach the catalytic surface, the adsorption-limited regime. The quality reduces with increasing argon partial pressure. As expected, this effect is larger with a lower ethylene flow and partial pressure.
The influence of hydrogen on the growth is largely attributed to its ability to reduce the catalyst and its presence in the exponential “start-up” term. However, the quality of the growth shows a dependence on the partial pressure of hydrogen as well. The existence of start-up effects are confirmed by comparing the results of experiments with two different runtimes. The influence of hydrogen on the start-up phase of the process is further confirmed by investigating the nucleation rate of the process. Finally, the reaction rate order of hydrogen is confirmed to be -1.

## Methods

### Growth of carbon nanotubes

The experimental setup consists of a feedback-controlled laser-assisted carbon nanotube growth process. This setup consists of an 808 nm CW-diode laser of maximum 30W, focused on a silicon substrate covered with 20 nm $Al_2O_3$ and a 1.5 nm Fe catalyst layer both deposited by e-beam evaporation. The laser is incident at an angle of 35° and has a spot-size of 800 µm. The reaction chamber is a stainless steel miniaturized reaction chamber where a laminar flow of process gases is controllably flowed over the substrate. The chamber is depicted in Fig 17. Analysis is carried out using a scanning electron microscope (FEI Quanta 600F ESEM) and a Raman spectrometer (632 nm, Horiba LabRAM HR).

The flow rate of one of each of the process gases is varied in a certain range, while maintaining the other two at a constant value using Bronkhorst Gas Mass Flow controllers. The basic configuration of gas flow rates was 200, 25 and 50 sccm, for argon, ethylene and hydrogen respectively while the exit pressure was 1 bar. This results in a partial pressure of 0.73 bar for argon, 0.09 bar for ethylene and 0.18 bar for hydrogen.

### Temperature evaluation

The growth process is controlled by monitoring the emitted thermal IR radiation, a first approximation of the temperature, using an InGaAs photodiode (1.2 – 2.6 µm range) that is placed beneath the substrate. A PID controller controls the temperature at the laser spot using Simulink.[13,14]

A detailed Finite Element Method model is developed using COMSOL Multiphysics 4.3a. The model combines and couples the heat and flow problem of the particular experiments.[13] The gas flow described in the paper can be varied in the model and results in a specific temperature evolution. The solids and gases present in the model incorporate temperature-dependent properties. Heat capacity, thermal conductivity and thermal expansion for solids; heat capacity, thermal conductivity, and thermal diffusivity for gases. For gases the ideal gas law is assumed. A prescribed mass flow is used for the inlet while the outlet is a laminar outflow - pressure. Further, on all exterior and interior free boundaries surface radiation is applied and natural convection in air is applied on the exterior free boundaries. The laser is modelled with a Gaussian intensity profile and applied as a boundary heat source. The recorded laser irradiance in time is used as the input. A K-type thermocouple placed at the edge of the substrate is used to calibrate and compare the simulations with experimental results.

### Growth rate and nucleation rate

To investigate growth kinetics, the reflected laser signal is dynamically monitored by a Si-detector. A sudden drop in reflection is attributed to the initiation and growth of the carbon nanotube forest. In general, a forest growth causes a decrease in reflection of 0.1 – 10% of the maximum value. To compare growth rates between experiments the maximum length of the nanotube forest is measured for different experiments terminated at the same value in reflection (in this case at 20% of the maximum value) and is divided by the time between nucleation ($t_n$) and passing of this value ($t_{20\%}$). For the surface diffusion regime, we have calculated the growth rate by a drop in reflected laser signal which was to about 60% of the maximum value, instead of the usual 20%, since the reflection in this regime did not drop any lower. In that case, the length that was used for the calculation was lower.

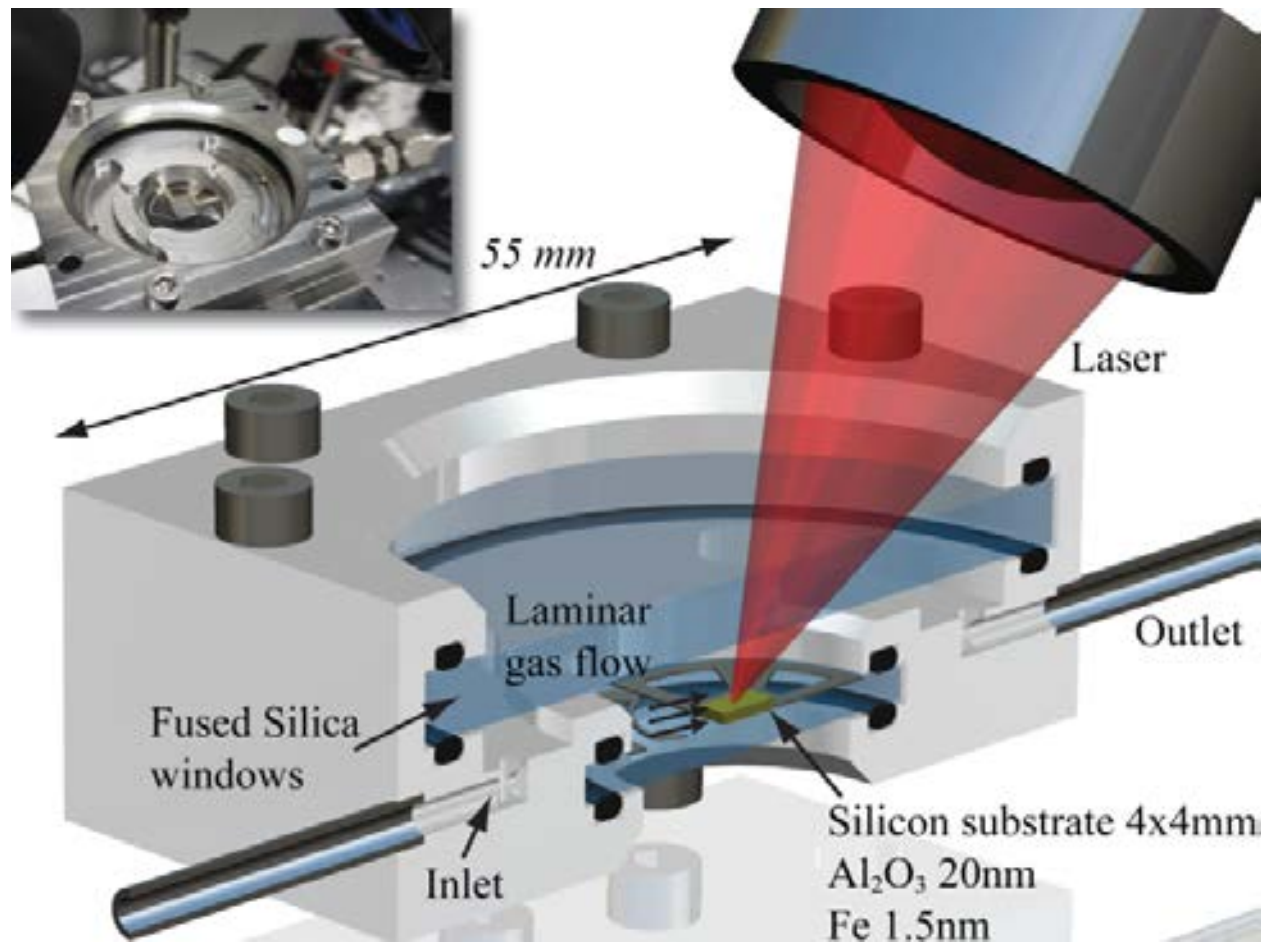

Fig 17. Schematic representation of the Laser-assisted CVD growth setup used in this work. Indicated are several components. The inset shows an actual photo of the setup.

The absorption coefficient $\alpha$ was calculated for different CNT lengths by measuring the intensity of the reflected and the diffused laser radiation. The latter was done by measuring the full spectrum intensity using a Tristan-USB UV/VIS/NEAR spectrometer.

We distinguish two types of experiments. The process can either be controlled to a set-point temperature after which the ethylene gas flow is added, or all three gases can be present before the process is started. The former is used to investigate the influence of temperature on the quality, morphology and growth rate of the CNT, while the latter is used to investigate nucleation characteristics.

## References

[a] Department of Mechanical Engineering, Eindhoven University of Technology, Den Dolech 2, Eindhoven, The Netherlands. E-mail: y.b.v.d.burgt@tue.nl

[b] Holst Centre/TNO – Netherlands Organization for Applied Scientific Research, HTC31, Eindhoven, The Netherlands.